\documentclass[a4paper]{vow2008}
\usepackage{graphicx}
\usepackage[english]{babel}
\usepackage[utf8]{inputenc}
\usepackage{natbib}

\newcommand{\urlnote}[1]{\footnote{\texttt{#1}}}
\newcommand{\ucdurl}[0]
{http://www.ivoa.net/Documents/latest/ UCDlist.html}
\newcommand{\ivoasemanticswgurl}[0]
{http://www.ivoa.net/cgi-bin/twiki/bin/view/ IVOA/IvoaSemantics}

\title{Data models for Radio Astronomy in the VO}
\author[1,2]{J.D. Santander-Vela}
\affil[1]{Instituto de Astrofísica de Andalucía-CSIC}
\affil[2]{AMIGA collaboration}

\begin{document}


	\maketitle

	\begin{abstract}
		Data Models are an essential part of automatic data processing,
		but even more so when trying to tie together data coming from
		many different data sources, as is the case for the
		International Virtual Observatory. In this talk we will review
		the different data models used in the IVOA, which parts of that
		Data Modelling work are still incomplete, specially in radio
		wavelengths, and the work the AMIGA group has done within the
		IVOA Data Modelling Working Group to overcome those
		shortcomings both in missing data models and support for Radio
		Astronomy.
	\end{abstract}

	\section{Introduction} 
	\label{sec:introduction}
	
	The AMIGA project (Analysing the interstellar Medium of Isolated
	GAlaxies) was born in 2003, and intends to provide a statistical
	characterisation of a strictly selected sample of isolated
	galaxies composed by more than 1000 objects, by means of
	multi-wavelength data, and with a particular emphasis on radio
	data at cm, mm, and sub-mm wavelengths. All these data are being
	periodically released via the web page of the
	project\footnote{\texttt{http://amiga.iaa.csic.es/}},
	which provides a Virtual Observatory (VO) ConeSearch interface.

	AMIGA+ is the natural extension to AMIGA, with three different
	goals: exploitation of the AMIGA catalogue, selecting the best
	candidates for a detailed study of isolated galaxies; scientific
	extension to the millimetre and submillimetre range; and
	participation in the development of systems allowing the access
	and display of large radio astronomical databases, both
	single-dish and interferometric, within the VO framework.
	
	During the AMIGA+ projects two VO-compliant radio astronomical
	archives were developed: the IRAM 30m antenna archive (soon to
	be published), and the DSS-63 archive. 

	Early during the development phase, a decision was made that we
	would not just provide a VO compatibility layer on top of these
	archives' infrastructures, but that the internal archive
	organisation should reflect existing VO data models in order to
	assure that the VO interface would be able to provide the most
	metadata.
	

	\section{Data models} 
	\label{sec:data_models}

	Data models are the detailed description of the set 
	of entities needed for information storage in a particular 
	field, and specify both the data being stored, and the
	relationships between them.
	Data models are part of the hidden VO infrastructure astronomers
	would normally never be involved with, but knowledge of data models
	can enhance the opportunities for the exploitation of the VO.

	Within the VO, data models apply not only directly to the
	scientific data, but to the metadata describing them. As the
	way to structure information depends on the application domain,
	VO data models describe astronomical datasets
	in a way that is as instrument independent as possible,
	to ensure that the
	same description can be used for data with different
	provenance. Users must also be able to query those data models
	to be able to find datasets which comply with certain properties.
	
	The IVOA Data Modelling Working Group (DMWG) started an
	an effort to provide a complete data model for astronomical
	observations, the \emph{Data Model for
	Observations}~\citep{2005dmo..rept.....M}. One of the most 
	important parts of it was the Characterisation of datasets, that
	is,
	the complete specification of where those datasets could be
	found in the spatial, temporal, and spectral axes, with more
	axes avalaible (i.e., polarisation) for suitable datasets.
	The Observation data model was put on hold, and the \emph{Data
	Model for Astronomical Dataset
	Characterisation}~\citep{McDowell:2007ly} was started.
	
	We have built a complete observation data model for single-dish
	radio telescopes, the Radio Astronomical DAta Model for Single-dish
	telescopes (RADAMS)~\citep{2008arXiv0810.0385S}, based on those
	two documents.

	
	\section{Data model elements} 
	\label{sec:data_model_elements}
	
	When defining a VO data model, we have to specify:
	
	\begin{description}
		\item[Entities] Being the data model building blocks,
		they group
		related attributes within a data model. They can be mapped to
		Classes in
		Object-Oriented Programming (OOP), or Elements in XML.
		
		\item[Fields] They are the actual data elements of the model.
		They map to Attributes in OOP, and they can be mapped to
		Attributes or to Elements without children in XML.
		
		\item[Relationships] The different entities and fields have
		hierarchical or relational relationships: an observation
		projects has projected observations, and all entities which
		share a common project ID are related, for instance. For the
		data model to be uniquely defined those relationships must be
		made explicit.
		
		\item[Data types] For computers to be able to correctly
		interpret a data stream a Data type needs to be specified.
		For instance, object IDs could be Integers, but they are
		normally textual, so String data must be used. We could
		consider the restrictions which can be defined for complex data
		types in XML as part of the data typing.
		
		\item[Units] No physical quantity can be specified without
		providing its units. Physical-data related Fields need Units to
		be specified, or Units have to be a fixed property
		of certain Fields, but they
		either need to exist as an implicit attribute of a particular
		field, or to have their own dedicated Field.
		
		\item[Semantics] As observation metadata are related to
		real-world elements and quantities, VO data models should
		specify semantics ---i.e., what is exactly
		meant in the real world by
		a particular field--- to avoid ambiguities. Most of VO
		semantics are provided via Unified Content Descriptors (UCDs)
		and UTypes.
	\end{description}

	
	\section{Semantics, UCDs, UTypes and IVOA vocabularies} 
	\label{sub:ucds_and_utypes}
		UCDs are a controlled vocabulary\urlnote{\ucdurl}, under
		the supervision of the IVOA Semantics WG, which provides a
		list of \emph{atoms} which can be used to identify fields
		as corresponding to specific astronomical quantities. For
		instance, a field containing the Right Ascension can be
		identified by the UCD atom \texttt{pos.eq.ra}, while a
		photometric flux in the V band can be identified by the two
		UCD atoms \texttt{phot.flux; em.opt.V}. This provides both
		a unified vocabulary to identify any astrophysical
		quantity, and an automatic knowledge discovery tool for
		fields with arbitrary relationships. In fact, UCDs were
		born out of a joint CDS/ESO data mining
		effort~\citep{1999ASPC..172..379O}.
		
		However, UCDs can only provide \emph{data kind}
		information, but not relationship information. In a sense,
		they are a kind of specialised \emph{unit}, complementary
		---orthogonal---
		to physical units: in the
		same way that quantities with the same physical units can
		be very different in nature (i.e., decay time for an
		isotope versus oscillation period), fields with identical
		UCDs can also be related to different real-word phenomena.
		In order to allow such deeper relationships to be
		expressed, and disambiguate metadata fields UTypes were
		born.
		
		UTypes are created from a hierarchical data model by 
		enumerating the different parents a particular field has
		in that hierarchy. For instance, a field containing the
		Right Ascension in equatorial coordinates for where an
		instrument was pointed to corresponds to the
		spatial coverage characterisation, in particular to the
		Location property, and thus it would sport a UType of
		\texttt{characterisation.coverage.spa\-tial.lo\-ca\-tion},
		the UCD would be \texttt{pos.eq.ra}, and its units
		could be any angular unit.
		
		But even with the help of units, UCDs and UTypes, sometimes it
		can be difficult to tag a particular piece of data with
		meaningful semantics, specially for data which does not have a
		direct place in a VO data model. For that we can borrow
		techniques from the Semantic Web (an effort for providing web
		documents with semantics, so that, for instance, a table of
		camera prices can be tagged so that software tools can identify
		in it prices, if possible belonging to digital cameras, even to
		particular brands), and provide one or more standardised
		astronomical vocabularies. The IVOA Semantics
		WG\urlnote{\ivoasemanticswgurl} has started recreating
		controlled vocabularies such as UCDs in Semantic Web form,
		and even the IAU thesaurus has been recreated in that
		way~\citep{Derriere:2008jo}. We are
		using them in the IRAM 30m archive in order to provide
		semantics to data coming from antenna engineering terms.
		

	\begin{figure*}[tbh]
		\centering
			\includegraphics[width=\textwidth]
			{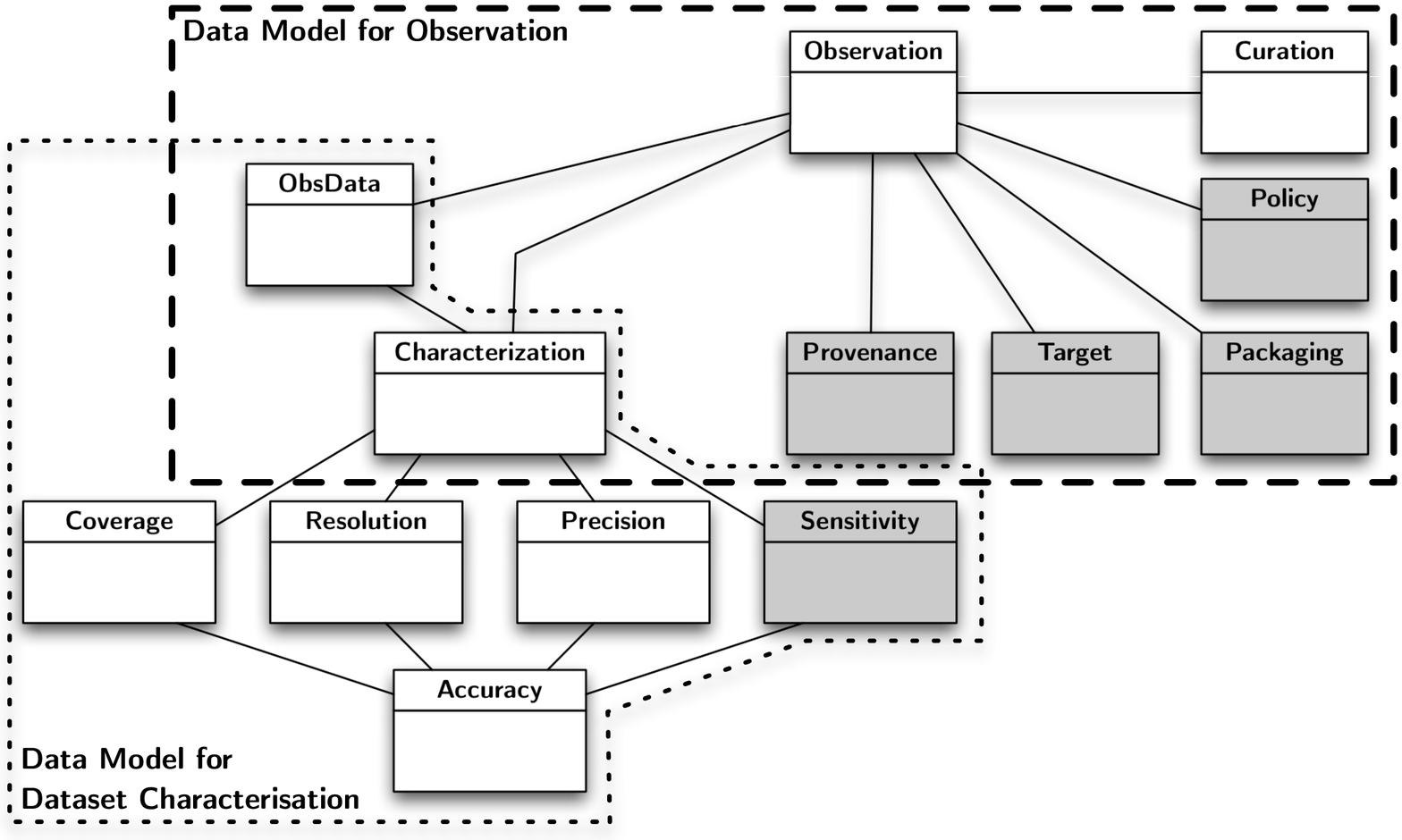}
		\caption{RADAMS structure: is a combination of the Observation
		and Characterisation data models, which have been completely
		specified for single dish radio telescopes. Data models which
		are completely new in their definition for any kind of
		instrument are marked in gray.}
		\label{fig:RADAMS}
	\end{figure*}

	\section{Role of data models in the VO} 
	\label{sec:role_of_data_models_in_the_vo}
	
	We can identify in the VO four different phases, and we can see
	that in all of them data models play a central role:
	
	\begin{description}
		\item[Discovery] Datasets available in the VO have to be
		discoverable for them to appear automatically in VO tools.
		The VO Registry holds data for existing datasets so that
		they can be easily discovered. The data model for
		Space-Time Coordinates (STC), dataset Characterisation
		(CharDM), Resource metadata (ResDM), and the UCD and IVOA
		thesaurus (IVOAT) are relevant in this phase.
		
		\item[Evaluation] Datasets have to be evaluated in order to
		assess their applicability to the kind of analysis we might
		wish to perform; for instance, in order to do image mosaicing
		we need a certain coordinate overlap, and in order to do
		image stacking we need an almost complete overlap, and
		comparable resolutions. The main data model involved in this
		phase is the CharDM.
		
		\item[Data Access] There is an implicit data model in the
		IVOA data access protocols, the Data Access Layer, which is
		centred on targets (coordinates with tolerances/search
		radii), and uses several properties from the CharDM, such as
		the Coverage in several axes.
		
		\item[Transformation] When creating a new dataset, or
		transforming an existing one, a new CharDM instance needs to be
		created. If the transformed data set is a spectrum, the
		Spectral data model (SpecDM) is needed both for obtaining the
		complete description of the original data and describing the
		transformed product. There is no existing data model yet for
		images or for more complex data within the VO. In addition, in
		order to trace the origin of the transformed image we would
		need to use a Provenance data model, that apart from being an
		integral part of the Observation data model (ObsDM), it should
		be built in a stand-alone form so that it can be applied to
		newly generated, non-observational data.
	\end{description}
	

	\section{RADAMS Structure and Properties} 
	\label{sec:radams_structure_and_properties}
	
	After having presented the importance of VO data models for the
	different activities involved in VO data queries, analyses and
	transformations, we will present the main RADAMS features.

	Figure~\ref{fig:RADAMS} shows the RADAMS main structure, which
	can be seen as a combination of the Observation data model and
	the CharDM data model, but fully specifying classes that were
	only laid out by the ObsDM.
	Some particular
	adaptations of the CharDM to radio astronomy,
	specifically for the Sensitivity class, have been performed, and
	the Target class is able to deal with radio catalogues. 
	
	However, RADAMS main contributions come from 
	the ObsDM classes which have been completely specified:
	
	\begin{description}
		\item[Packaging] We have developed a VOPack packaging standard,
		which embeds characterisation information for any packaged
		observation set, and can specify the recursive inclusion of
		other packages.
		
		\item[Policy] The Policy class that we have developed is able
		to accommodate different user roles determined from the user
		identification and the explicit or implicit policies for each
		particular dataset.
		
		\item[Provenance] Initially specified only for radio astronomy
		data, we are working to make it more general, and to be able
		to use it outside of observational scopes,
		so that software tools which provide dataset transformations
		can document the origin of their processed datasets.
	\end{description}
	
	The RADAMS has been implemented in two archives: the Telescope
	Access for Public Archive System (TAPAS) for the IRAM 30m, to
	be announced in the February 2009 IRAM Newsletter, and the
	scientific archive of NASA's Deeps Space Center in Madrid 70~m
	antenna, DSS-63, and has shown that VO data models can be used as
	a blueprint for archives which are being built from scratch for
	VO compatibility.
	

	\section{Future work} 
	\label{sec:future_work}
	
	In the future, we plan to have the RADAMS Provenance data model,
	part of the author's thesis (to be published in 2009), contributed
	to the IVOA DMWG, and have it integrated with the ObsDM.
	
	We are starting also a collaboration with the ALMA Scientific
	Archive team in order to create in the future both a data model
	for ALMA data cubes which can be integrated in a future IVOA
	data model for multidimensional data, and providing VO
	compatibility to the ALMA Science Archive.
	
	
	\section{Conclusions} 
	\label{sec:conclusions}

		Data models are one of the three bases interoperability relies
		on. As the data models for astrophysical observations' metadata
		must be interoperable across different software packages and
		instrument domains, not every detail can be included in the
		data models, or can be described using IVOA's UCDs and UTypes.
		Semantic web technologies, such as thesauri expressed in 
		W3C standards, can be used to provide additional semantics.

		However, the IVOA high-level data modelling efforts are
		complete enough as to have guided the development of the RADAMS
		data model. The RADAMS has been successfully used to create two
		operational radio astronomical archives for two very different
		antennas and instrument systems, such as the IRAM 30m and the
		DSS-63.
		
		In the future, the ObsDM must be complete enough to support
		the complex datasets to be produced by ALMA and radio
		telescopes further ahead in time, such as the LOFAR and other
		SKA pathfinders. 
	
	\section*{Aknowledgments} 
	\label{sec:aknowledgments}
		The author wishes to acknowledge Emilio García for his support
		in the early development of the RADAMS, his discussions with
		Stephane Leon for clarifications and enrichment of the RADAMS,
		and the direction of José Francisco Gómez and Lourdes
		Verdes-Montenegro during my Master Thesis, were most of this
		work was developed. Further discussions with J.~E.~Ruiz del
		Mazo and V.~Espigares during the implementation phase are also
		acknowledged, as well as the discussions inside the IVOA Data
		Modeling Working Group. Part of this work has been performed
		with the funding of Spanish Education Ministry projects AYA
		2005-07516-C02 and AYA2005-24102-E, and the European FP7
		EuroVO-AIDA project.
		

	\bibliographystyle{vow2008}
	\bibliography{DataModel4RadioAstronomy}

\end{document}